\documentclass{svjour2}                    
\smartqed  
\usepackage{graphicx}
\journalname{Journal of Superconductivity and Novel Magnetism}
\begin{document}

\title{Single crystal growth of BaFe$_{2-x}$Co$_x$As$_2$ without fluxing agent}

\author{Changjin Zhang         \and
        Lei Zhang        \and
        Chuanying Xi        \and
        Langsheng Ling        \and
        Wei Tong        \and
        Shun Tan        \and
        Yuheng Zhang        \and
}

\institute{Changjin Zhang        \and
        Lei Zhang        \and
        Langsheng Ling        \and
        Wei Tong        \and
        Yuheng Zhang     \at
              High Magnetic Field Laboratory, Chinese Academy of Sciences, Hefei 230031, People's Republic of China \\
              Tel.: +86-551-559-1672\\
              Fax: +86-551-559-1149\\
              \email{zhangcj@hmfl.ac.cn}           
           \and
        Shun Tan        \and
        Yuheng Zhang     \at
              Hefei National Laboratory for Physical Sciences at Microscale, University
of Science and Technology of China, Hefei 230026, People's Republic
of China }

\date{Received: date / Accepted: date}

\maketitle

\begin{abstract}
We report a simple, reliable method to grow high quality
BaFe$_{2-x}$Co$_x$As$_2$ single crystal samples without using any
fluxing agent. The starting materials for the single crystal growth
come from well-crystallized polycrystalline samples and the highest
growing temperature can be 1230$^\circ$C. The as-grown crystals have
typical dimensions of 4$\times$3$\times$0.5 mm$^3$ with $c$-axis
perpendicular to the shining surface. We find that the samples have
very large current carrying ability, indicating that the samples
have good potential technological applications. 
\end{abstract}

\section{Introduction}
\label{intro}

The discovery of superconductivity in iron-pnictide systems has
attracted tremendous interests not only due to its scientific value
but also its potential industrial applications [1]. The relatively
high transition temperature, highly flexibility, very high
upper-critical magnetic field and other physical qualities make the
iron-pnictide systems very useful in industry [2-4]. The existing
challenges, such as optimizing synthesis methods for technological
applications and clarifying the ambiguity in the superconducting
mechanism, will keep iron-pnictide systems on the frontiers of
research for a long time, in parallel to high-$T_c$ cuprates [4].

In order to determine the application parameters which are important
to commercial use, great efforts have been made to grow high-quality
single crystals of iron-pnictide superconductors [5-8] A lot of
physical properties, such as the transition temperature, the upper
critical field, the vortex structure, etc., have been determined
using single crystal samples. However, due to the relatively high
melting temperature of the iron-pnictide samples, the single
crystals of iron-pnictides are generally grown by using self-flux
method or flux method where excess FeAs mixture or Sn is used as the
fluxing agent. The advantage of these methods is that the melting
temperature can be significantly decreased comparing to the melting
point of the crystal itself. While the disadvantage of these methods
is unnegligible. For example, if we grow BaFe$_{2-x}$Co$_x$As$_2$
single samples using excess Fe(Co)As mixture as fluxing agent, the
actual Fe/Co ratio can not be accurately controlled in the growth
procedure. If one uses Sn as fluxing agent, the problem is that one
can not remove the Sn from the surface of the sample easily [5]. In
this paper we report a simple, reliable method to grow high quality
BaFe$_{2-x}$Co$_x$As$_2$ single crystal samples without using any
fluxing agent. The samples have typical dimensions of
4$\times$3$\times$0.5mm$^3$ with $c$-axis perpendicular to the
shining surface. The critical current density of the samples are
also determined. The critical current density without external
magnetic field is quite high, meaning large current carrying ability
of the samples, which points to optimistic applications.

\section{Experimental detail}
Single crystal samples were grown using well-crystalized
BaFe$_{2-x}$Co$_x$As$_2$ polycrystalline samples as the starting
materials. The polycrystalline samples with nominal composition
BaFe$_{2-x}$Co$_x$As$_2$ were prepared by conventional solid-state
reaction method using high-purity Ba (crystalline dendritic solid,
99.9\%, Alfa-Aesar), Fe (powder, 99.9\%, Alfa-Aesar), Co (powder,
99.9\%, Alfa-Aesar), and As (powder, 99\%, Alfa-Aesar) as starting
materials. The crystalline dendritic solid Ba was pressed into thin
pellet using an agate mortar and was cut into very small size
(typically less than 0.5$\times$0.5 mm$^2$). The raw materials were
mixed and wrapped up by Ta foil and sealed in an evacuated quartz
tube. They were pre-heated at 600$^{\circ}$C for 12 hours and cooled
down slowly to room temperature. The mixture was then ground and
pressed into pellets and heated at 900$^{\circ}$C for 24 hours. When
the furnace was cooled down, the pellets were taken out and placed
in an argon-filled glove box. We performed powder x-ray diffraction
measurements on these samples and found that the samples were all in
single phase.

The polycrystalline powder was pressed into pellets and placed in a
quartz tube in an Argon-filled glove box. The quartz tube was sealed
after it was evacuated by a molecular pump. Then the quartz tube was
placed into a box furnace. The furnace was heated to 1230$^\circ$C
at a rate of 60$^\circ$C per hour. After holding at 1230$^\circ$C
for 12 hours, it was cooled to 850$^\circ$C at 2$^\circ$C per hour
followed by furnace cooling to room temperature. The quartz tube was
found almost intact after the whole procedure. When we break the
quartz tube and pick out the sample, slides of samples with shining
surfaces can be easily cleaved. It should be noted that we have
tried to melt the samples at even higher temperature using a
double-wall quartz tube. However, we find that the samples begin to
decompose at temperature higher than 1240$^\circ$C.

X-ray diffraction (XRD) was carried out by a Rigaku-D/max-gA
diffractometer using high-intensity Cu-K$\alpha $ radiation to
screen for the presence of an impurity phase and the changes in
structure. The homogeneity and chemical compositions of the samples
were examined using an energy dispersive x-ray spectrometer (EDXS).
The resistivity was measured using a standard four-probe method in a
closed-cycle helium cryostat. The magnetic susceptibility and the
magnetic hysteresis loops of the samples were determined by a SQUID
magnetometer (Quantum Design, MPMS).

\section{Results and discussion}

\begin{figure}
  \includegraphics[width=0.75\textwidth]{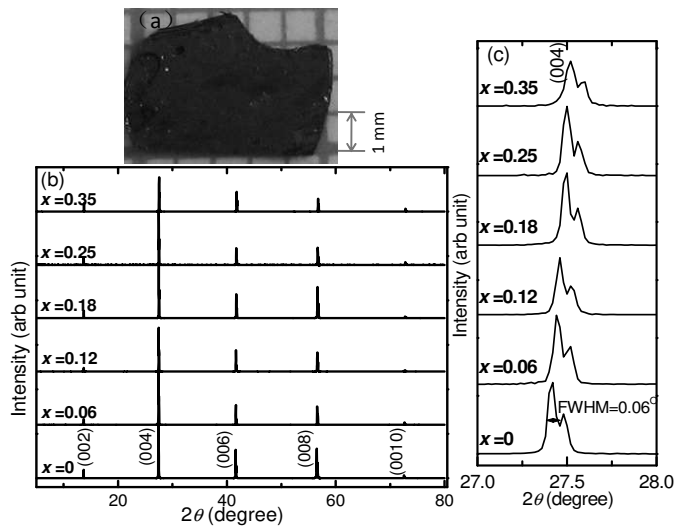}
\caption{(a) Photograph of a as-grown BaFe$_2$As$_2$ single crystal.
(b) X-ray diffraction pattern at room temperature for the
BaFe$_{2-x}$Co$_x$As$_2$ single crystals. (c) An enlarge view of the
(004) reflection.}
\label{fig1}       
\end{figure}

Figure 1(a) shows a picture of a single crystal sample which has
dimensions of about 4.5$\times$3$\times$0.5mm$^3$. We select several
pieces of crystal and perform EDXS measurement and find that the
Co-contents in all pieces are close to the nominal compositions,
indicating that the samples having shining surface are chemically
homogeneous. The nominal and measured compositions of the selected
samples are summarized in table 1. In order to judge the orientation
of the samples, we perform x-ray diffraction (XRD) measurement on
the as-grown samples. Figure 1(b) gives the typical XRD patterns of
the BaFe$_{2-x}$Co$_x$As$_2$ ($x$=0, 0.06, 0.12, 0.18, 0.25, and
0.35) samples. Only the (00$l$) diffraction peaks with even $l$ are
observed, confirming that the crystallographic $c$-axis is
perpendicular to the shining surface. For all the diffraction peaks,
the full width at half maximum (FWHM) is less than 0.06$^\circ$,
indicating the excellent quality of the single crystals. In order to
see the shift of the peaks clearly, we plot in Fig. 1(c) the
enlarged view of the (004) reflection. One can see that all the
reflections are splitted into two shoulder peaks. The shoulder peak
at lower angle is the reflection of the Cu-K$\alpha$$_1$ radiation
and the one at higher angle is the reflection of the
Cu-K$\alpha$$_2$ radiation. It can be seen that the (004) peak
slightly shifts to higher angle with increasing Co content, meaning
that the $c$-axis constant decreases monotonously as the Co content
is increased. The calculated $c$-axis lattice contents for the
samples are given in Table 1.

\begin{figure}
  \includegraphics[width=0.75\textwidth]{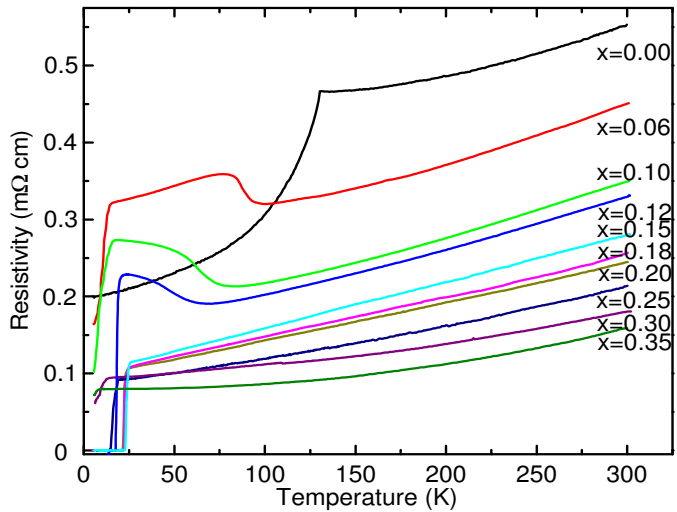}
\caption{Temperature dependence of in-plane resistivity
for the BaFe$_{2-x}$Co$_x$As$_2$ samples.}
\label{fig2}       
\end{figure}

The superconducting properties of the BaFe$_{2-x}$Co$_x$As$_2$
single crystals are given in Fig. 2. The superconductivity emerges
in the $x$$\geq$0.06 samples. And the maximum critical transition
temperature $T_{c,\rho=0}$ reaches to 23.3 K at the optimal doping
concentration $x$=0.15. With further increasing Co doping content,
$T_c$ decreases monotonously. The superconductivity disappears when
$x$$>$0.35.

\begin{figure}
  \includegraphics[width=0.75\textwidth]{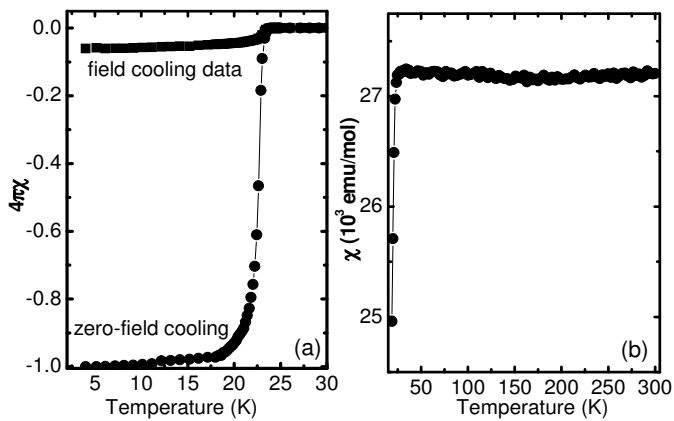}
\caption{(a) Temperature dependence of magnetic susceptibility for
BaFe$_{1.80}$Co$_{0.20}$As$_2$ both under zero-field cooling
condition and under field-cooling condition at 10 Oe. (b) The
magnetization as the function of temperature under 1 Tesla magnetic
field.}
\label{fig3}       
\end{figure}

Figure 3(a) gives the temperature dependence of magnetic
susceptibility below $T_c$ for the $x$=0.20 sample both under
zero-field cooling condition and under field-cooling condition at 10
Oe. It is found that the superconducting transition occurs at 23.1
K, consistent with the resistivity results. For the magnetic
susceptibility at $T$$>$$T_c$, the susceptibility signal is almost
undetectable within the accuracy limit of the $Quantum$ $Design$
MPMS magnetometer (about 10$^{-8}$ emu). In order to know the
magnetic state at the normal state, we measure the temperature
dependence of magnetic susceptibility under 1 Tesla. The result is
shown in Fig. 3(b). From Fig. 3(b) we notice that the magnetic
susceptibility exhibits almost temperature-independent behavior
above $T_c$, indicating that the magnetic state of the
BaFe$_{1.80}$Co$_{0.20}$As$_2$ system can not be the Curie
paramagnetism. The fact that the magnetization is very weak and
temperature-independent suggest that the paramagnetic state is a
Pauli-paramagnetic state, which is consistent with the metallic
behavior of the BaFe$_{1.80}$Co$_{0.20}$As$_2$ system. The
predominant Pauli-paramagnetic state in the Co-doped sample suggest
that the magnetic moment of the electrons near the Fermi surface
should be delocalized. Previous neutron scattering experiments on
CaFe$_2$As$_2$ have suggested that the magnetism is neither purely
local nor purely itinerant and that it is a complicated mix of the
two [9]. Here the predominant Pauli-paramagnetic state in the
Co-doped sample suggest that the itinerant moments might be dominate
in the superconducting sample.

\begin{figure}
  \includegraphics[width=0.75\textwidth]{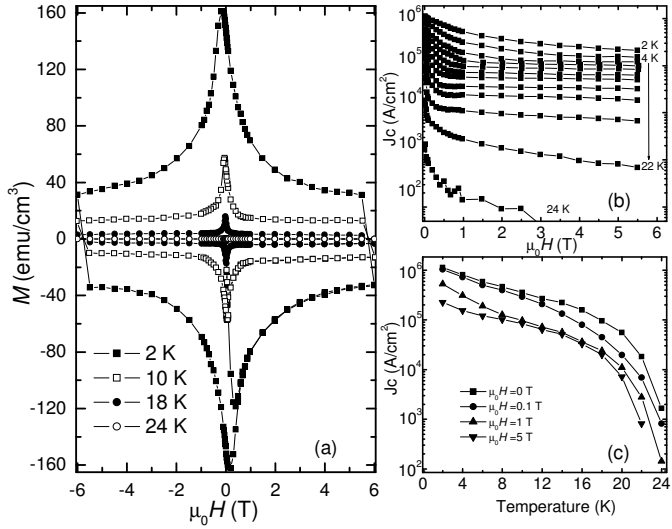}
\caption{The magnetization as the function of external magnetic
field below $T_c$ for the $x$=0.20 sample. (b) The critical current
density as the function of magnetic field for the $x$=0.20 sample at
different temperatures. (c) The temperature dependence of critical
current density for the $x$=0.20 sample under zero-field and under
external magnetic field.}
\label{fig4}       
\end{figure}

Figure 4(a) shows magnetic hysteresis loops at various temperatures
below $T_c$ calculated by applying the magnetic field up to 6 T. The
$M$$\sim$$H$ curves exhibit a central peak at zero magnetic field
and the magnetization decreases continuously with increasing
magnetic fields. The sharp peak around $\mu$$_0$$H$ = 0 is similarly
observed in other iron-pnictide materials [8, 10-11]. Figure 4(b)
shows the magnetic field dependence of the critical current density
$J_c$ derived from the hysteresis loop width by Bean critical state
model using the relation $J_c$ = 20$\triangle$$M$$/$$a$(1- $a/3b$)
[12], where $a$ and $b$ are the width and length of the sample,
respectively ($a$$<$$b$), and $\triangle$$M$ is the difference
between the upper and the lower branches in the $M$$\sim$$H$ loops.
It is found from Fig. 4(b) that the critical current density $J_c$
of the sample reaches to 1.2$\times$10$^6$ A/cm$^2$ without external
magnetic field. We notice that this $J_c$ value is higher than
previous reported $J_c$ value of BaFe$_{2-x}$Co$_x$As$_2$ single
crystal samples, either grown using self-flux method or using flux
method [8,13-15]. For example, the $J_c$ values of recent grown
Co-doped BaFe$_2$As$_2$ single crystal thin films are within the
range of 60-100 kA/cm$^2$ at 12 K (without external magnetic field)
[8], which is less than the value of 280 kA/cm$^2$ in present
sample. The $J_c$ value of a BaFe$_{1.80}$Co$_{0.20}$As$_2$ sample
grown by self-flux method is about 6$\times$10$^5$ A/cm$^2$ at 5 K
[13]. For a BaFe$_{1.852}$Co$_{0.148}$As$_2$ single crystal grown
using Sn flux, the $J_c$ value at 16 K under 6 Tesla is about 5
kA/cm$^2$ [14], which is also less than the value of 26 kA/cm$^2$ in
present case. Based on these facts we suggest that the samples grown
without any fluxing agent may have better current carrying ability
comparing to those from flux growth. But this value is slightly less
than the highest critical current density of 4 MA/cm$^2$ in Co-Doped
BaFe$_2$As$_2$ epitaxial films which was recently grown on
(La,Sr)(Al,Ta)O$_3$ substrates [16]. The comparison between single
crystals grown using different methods reveals that further
improvement of critical current density is still possible.
Considering that the BaFe$_{2-x}$Co$_x$As$_2$ samples have upper
critical field as high as 60 T, critical temperatures of above 20 K,
low anisotropy, and, as shown here, high intrinsic critical current
density, these materials can be considered as good candidates for
applications.

The $J_c$ value decreases both with increasing temperature and with
increasing external magnetic field, as can be seen from Figs. 4(b)
and (c). At low temperatures ($\leq$20 K), the trend of $J_c$ decay
is similar to that of conventional high-$T_c$ cuprates [17]. At high
temperature ($>$20 K), the flux creep effect is evident by showing
relatively strong dependence of critical current density on the
external magnetic field [18].

\section{Conclusions}
In summary, we have grown large-size Co-doped BaFe$_2$As$_2$ single
crystals without using any fluxing agent. We find that the as-grown
samples have larger current carrying ability comparing to those
grown with the aid of fluxing agent, indicating promising industrial
applications.

\begin{table}
\caption{The comparison between nominal and real compositions and
the $c$-axis lattice parameters of the BaFe$_{2-x}$Co$_x$As$_2$
samples}
\label{tab:1}       
\begin{tabular}{lll}
\hline\noalign{\smallskip} Nominal composition & real composition &
$c$ (\AA) \\
\noalign{\smallskip}\hline\noalign{\smallskip}
BaFe$_2$As$_2$ & BaFe$_2$As$_2$ & 13.018(4) \\
BaFe$_{1.9}$Co$_{0.1}$As$_2$ & BaFe$_{1.9}$Co$_{0.1}$As$_2$ & 13.004(4) \\
BaFe$_{1.8}$Co$_{0.2}$As$_2$ & BaFe$_{1.81}$Co$_{0.19}$As$_2$ & 12.983(2) \\
BaFe$_{1.7}$Co$_{0.3}$As$_2$ & BaFe$_{1.71}$Co$_{0.29}$As$_2$ & 12.956(5) \\
\noalign{\smallskip}\hline
\end{tabular}
\end{table}

\begin{acknowledgements}
This work was supported by the State Key Project of Fundamental
Research of China through Grant 2010CB923403 and 2011CBA00111, and
the Hundred Talents Program of the Chinese Academy of Sciences.
\end{acknowledgements}


\begin{thebibliography}{}
%
%
\bibitem{} Y. Kamihara, T. Watanabe, M. Hirano, and H. Hosono, J. Am. Chem. Soc. 130, 3296 (2008).

\bibitem{} F. Hunte, J. Jaroszynski, A. Gurevich, D. C. Larbalestier, R. Jin, A. S. Sefat, M. A. McGuire, B. C. Sales,
D. K. Christen, and D. Mandrus, Nature 453, 903 (2008).

\bibitem{} K. Ishida, Y. Nakai, and H. Hosono, J. Phys. Soc. Jpn. 78, 062001 (2009).

\bibitem{} P. M. Aswathy, J. B. Anooja. P. M. Sarun, and U. Syamaprasad, Supercond. Sci. Technol. 23, 073001 (2010).

\bibitem{} N. Ni, S. L. Bud'ko, A. Kreyssig, S. Nandi, G. E. Rustan, A. I. Goldman, S. Gupta, J. D. Corbett,
A. Kracher, and P. C. Canfield, Phys. Rev. B 78, 014507 (2008).

\bibitem{} F. Hardy, P. Adelmann, T. Wolf, Hilbert v. Lohneysen, and C. Meingast, Phys. Rev. Lett. 102, 187004 (2009).

\bibitem{} J. S. Kim, E. G. Kim, and G. R. Stewart, J. Phys.: Condens. Matter 21, 252201 (2009).

\bibitem{} C. Tarantini, S. Lee, Y. Zhang, J. Jiang, C. W. Bark, J. D. Weiss, A. Polyanskii, C. T. Nelson, H. W. Jang,
C. M. Folkman, S. H. Baek, X. Q. Pan, A. Gurevich, E. E. Hellstrom,
C. B. Eom, and D. C. Larbalestier, Appl. Phys. Lett. 96, 142510
(2010).

\bibitem{} J. Zhao, D. T. Adroja, D.-X Yao, R. Bewley, S. L. Li, X. F. Wang, G. Wu, X. H. Chen, J. P. Hu, and P. C. Dai,
Nature Phys. 5, 555 (2009).

\bibitem{} R. Prozorov, M. A. Tanatar, N. Ni, A. Kreyssig, S. Nandi, S. L. Bud'ko, A. I. Goldman, and P. C. Canfield,
Phys. Rev. B 80, 174517 (2009).

\bibitem{} F. Kametani, P. Li, D. Abraimov, A. A. Polyanskii, A. Yamamoto, J. Jiang, E. E. Hellstrom, A. Gurevich,
D. C. Larbalestier, Z. A. Ren, J. Yang, X. L. Dong, W. Lu, Z. X.
Zhao, Appl. Phys. Lett. 95, 142502 (2009).

\bibitem{} C. P. Bean, Rev. Mod. Phys. 36, 31 (1964).

\bibitem{} Y. Nakajima, T. Taen, and T. Tamegai, J. Phys. Soc. Jpn. 78, 023702 (2009).

\bibitem{} M. A. Tanatar, N. Ni, S. L. Bud'ko, P. C. Canfield, and R. Prozorov, Supercond. Sci. Technol. 23, 054002 (2010).

\bibitem{} R. Prozorov, M. A. Tanatar, N. Ni, A. Kreyssig, S. Nandi, S. L. Bud'ko, A. I. Goldman, and P. C. Canfield,
Phys. Rev. B 80, 174517 (2009).

\bibitem{} T. Katase, H. Hiramatsu, T. Kamiya, and H. Hosono, Appl. Phys. Exp. 3, 063101 (2010).

\bibitem{} C. Cai, B. Holzapfel, J. H\"{a}nisch, L. Fern\'{a}ndez and L. Schultz, Phys. Rev. B 69, 104531 (2004).

\bibitem{} T. Matsushita, T.Fujiyoshi, K.Toko and K.Yamafuji, Appl. Phys. Lett. 56, 2039 (1990).

\end{thebibliography}


\bigskip

\bigskip

\end{document}